\def\papertitle{Evaluating Dynamic Range Compressor Models Using Control-Voltage Measurements: an Approach and Dataset}
\def\paperauthorA{Benjamin R. Thompson}
\def\paperauthorB{Michael C. Heilemann}

\documentclass[twoside,a4paper]{article}
\usepackage{etoolbox}

\usepackage[print]{dafx26v3}

\usepackage{amsmath,amssymb,amsfonts,amsthm}
\usepackage{siunitx}
\usepackage{euscript}
\usepackage[T1]{fontenc}
\usepackage[utf8]{inputenc}
\usepackage{ifpdf}
\usepackage[english]{babel}
\usepackage{caption}
\usepackage{subfig} 
\usepackage{color}
\usepackage{booktabs}
\usepackage{lipsum}
\usepackage{xspace}
\usepackage[table]{xcolor}

\newif\ifblind
\blindfalse 

\newcommand{\redact}[1]{%
\ifblind
[redacted for blind review]%
\else
#1
\fi
}

\newcommand{\redactcite}[1]{%
\ifblind
[citation omitted for blind review]%
\else
#1
\fi
}

\input glyphtounicode
\ninept

\newcounter{numauth}
\newcounter{listcnt}
\newcommand\authcnt[1]{\ifdefined#1 \stepcounter{numauth} \fi}

\newcommand\addauth[1]{
\ifdefined#1 
\stepcounter{listcnt}
\ifnum \value{listcnt}<\value{numauth}
\appto\authorslist{, #1}
\else
\appto\authorslist{~and~#1}
\fi
\fi}
\authcnt{\paperauthorB}
\authcnt{\paperauthorC}
\authcnt{\paperauthorD}
\authcnt{\paperauthorE}
\authcnt{\paperauthorF}
\authcnt{\paperauthorG}
\authcnt{\paperauthorH}
\authcnt{\paperauthorI}
\authcnt{\paperauthorJ}
\def\authorslist{\paperauthorA}
\addauth{\paperauthorB}
\addauth{\paperauthorC}
\addauth{\paperauthorD}
\addauth{\paperauthorE}
\addauth{\paperauthorF}
\addauth{\paperauthorG}
\addauth{\paperauthorH}
\addauth{\paperauthorI}
\addauth{\paperauthorJ}

\usepackage{times}

\newif\ifpdf
\ifx\pdfoutput\relax
\else
   \ifcase\pdfoutput
      \pdffalse
   \else
      \pdftrue
   \fi
\fi

\ifpdf 
  \usepackage[pdftex,
    pdftitle={\papertitle},
    pdfauthor={\authorslist},
    pdfsubject={Proceedings of the 29th International Conference on Digital Audio Effects (DAFx26)},
    colorlinks=false, 
    bookmarksnumbered, 
    pdfstartview=XYZ 
  ]{hyperref}
  \usepackage[pdftex]{graphicx}
\else 
  \usepackage[dvips]{epsfig,graphicx}
  \usepackage[dvips,
    pdftitle={\papertitle},
    pdfauthor={\authorslist},
    pdfsubject={Proceedings of the 29th International Conference on Digital Audio Effects (DAFx26)},
    colorlinks=false, 
    bookmarksnumbered, 
    pdfstartview=XYZ 
  ]{hyperref}
\fi
\usepackage[hypcap=true]{caption}
\title{\papertitle}

\affiliation
{\paperauthorA\ and \paperauthorB}
{\href{https://www.hajim.rochester.edu/ece/}{Dept. of Electrical \& Computer Engineering} \\ University of Rochester \\ Rochester, USA\\
{\tt \href{mailto:bthomp23@ur.rochester.edu}{bthomp23@ur.rochester.edu}}
}

\DeclareMathOperator*{\argmin}{arg\,min}

\begin{document}

\ifpdf 
  \DeclareGraphicsExtensions{.png,.jpg,.pdf}
\else  
  \DeclareGraphicsExtensions{.eps}
\fi


\maketitle

\begin{abstract}
The quantity that defines the behavior of a dynamic range compressor is the time-varying gain applied to the signal as a function of the input level. However, models of these devices are typically evaluated using proxy metrics because isolating the gain reduction signal from the audio input--output data included in existing datasets creates an ill-conditioned inverse problem. It is unclear how accurately these metrics describe the behavior the model is tasked with emulating, particularly as waveform-based metrics can be influenced by secondary effects introduced by analog processing and capture, even when those effects are inaudible. We investigate a method of evaluation in which the gain-reduction signal produced by a model is measured directly against a gain-reduction control voltage signal produced by the hardware. To evaluate the efficacy of this metric as a learning objective, a gray-box model is trained using loss computed directly over the gain control signals alongside two models trained using common proxy losses. The models trained using proxy losses did not achieve parity with models trained directly on the gain control signal when evaluated with respect to the underlying control trajectory, and the waveform-domain metrics assigned similar errors to models that were clearly separated by the direct metric. To facilitate further exploration of this method of evaluation, we present a Solid State Logic bus compressor dataset that includes the gain control voltage signal captured alongside the audio output.
\end{abstract}

\section{Introduction}
\label{sec:intro}
A dynamic range compressor (DRC) is a processor that applies a time-varying attenuation to a signal as a function of its level. In general, attenuation is only applied when the input level exceeds a threshold set by the operator, and the amount of attenuation is proportional to the amount by which the signal exceeds that threshold. The method used to calculate signal level, the mapping from signal level to attenuation, and the ballistics that govern the application of that attenuation are implementation-specific \cite{self2010audio}. As a result, in a creative context, certain DRCs are often preferred for the behaviors that arise from these design details, and it can be desirable to capture the behavior of a specific hardware implementation in a digital model.

The design and analysis of digital DRCs using traditional digital signal processing (DSP) techniques is well established \cite{giannoulis2012digital, dafx_ch4_2011, abel2003peak}. In recent years, neural approaches to modeling dynamic range compression have been widely explored. These include black-box methods that learn the input--output mapping directly from data \cite{hawley2019signaltrain, steinmetz2021efficient, simionato2023fully}, as well as gray-box approaches that combine neural networks with structured, differentiable DSP models \cite{engel2020ddsp, wright2022grey, yu2024differentiable}.

To support the training of black- and gray-box models and the validation of white-box models, a number of publicly available datasets for dynamic range compression have been proposed \cite{comunita2025differentiableblackboxgrayboxmodeling, comunita2023modelling, simionato2023fully}, including a recent large-scale collection of data processed by an SSL G-Comp bus compressor \cite{gu2025solidstatebuscomplargescale}. 

The ideal, discrete version of the relationship that defines a DRC is
\begin{equation}
    y_{\mathrm{dB}}[k]=x_{\mathrm{dB}}[k]+g_{\mathrm{dB}}[k],
\label{eq:main_dB}
\end{equation}
where $y_{\mathrm{dB}}$ is the output signal in decibels (\si{\decibel}), $x_{\mathrm{dB}}$ is the input signal in \si{\decibel}, $g_{\mathrm{dB}}$ is the gain control signal in \si{\decibel}, and $k$ is the sample index.

It follows from Eq.~\eqref{eq:main_dB} that the accuracy of any model of the ideal system is defined by the accuracy of the quantity $g_{\mathrm{dB}}[k]$. For a hardware ground truth, the signal has been modified by secondary effects, including phase shifts, amplitude-response deviations, and noise contamination, in addition to the intentional amplitude modulation imposed by $g_{\mathrm{dB}}[k]$. As such, extracting $g_{\mathrm{dB}}[k]$ represents an ill-conditioned inverse problem \cite{tikhonov1977solutions} that is further complicated by numerically unstable sample-wise division. As a result, proxy metrics operating over the output waveforms are employed to train and evaluate compressor models. These proxy metrics are sensitive to the same secondary effects that make gain extraction impractical, limiting their ability to reflect error in $g_{\mathrm{dB}}[k]$.

In each of the datasets referenced above, the signals provided are limited to audio input--output pairs where the underlying gain control signal is not directly observable. This work presents a dataset of music and calibration signals processed by an SSL Logic FX G384 bus compressor, in which each audio input–output pair is accompanied by the corresponding gain control signal~\redactcite{\cite{thompson_ssl_cv_dataset_2026}}. The inclusion of $g_{\mathrm{dB}}[k]$ in the dataset allows model performance to be measured directly against the quantity that defines the processor’s behavior and further enables proxy metrics themselves to be evaluated against a ground-truth error defined on the gain trajectory.

The SSL bus compressor was chosen both for its relevance in the audio industry, as evidenced by the number of commercial emulations available \cite{waves_ssl_g_master, ua_ssl_g_bus_comp, ssl_native_bus_comp_2}, and because aspects of its implementation are not captured by generic DRC models, including a curved static response and program-dependent release behavior, in which the effective release time is not fixed but instead depends on the recent history of the input signal.

The remainder of this paper is organized as follows. Two proxy metrics commonly used in compressor modeling are discussed in Section~\ref{sec:metrics}, and the use of control-signal error to evaluate the behavioral accuracy of such models is motivated; an experimental comparison of loss regimes is described in Section~\ref{sec:experiment}; and the proposed SSL bus compressor dataset and measurement procedure are presented in Section~\ref{sec:dataset}.

\section{Loss Functions and Evaluation Metrics for Compressor Models}
\label{sec:metrics}

The principal effect of dynamics processing is a time-varying modification of the signal amplitude. Hardware processors inevitably introduce secondary effects such as frequency-dependent changes in phase and amplitude response and the accumulation of additive noise. When developing a digital model for creative use, these secondary effects can generally be ignored where they are perceptually insignificant or deemed undesirable, or modeled separately where they are considered desirable. However, when evaluating the behavioral accuracy of a DRC model using metrics computed over the processed waveforms, secondary effects and capture artifacts can obscure the true error in the gain-reduction signal.

Consider the discrete outputs for a hardware compressor and a digital model of that compressor, $y[k]$ and $\hat{y}[k]$ respectively. In both cases, the discrete input signal is $x[k]$. A common metric used to quantify the accuracy of a compressor model is $L_1$, or Mean Absolute Error over the processed waveforms, computed as 
\begin{equation} 
    L_1 = \frac{1}{K}\sum_{k=1}^{K}\left\lvert  y[k] - \hat{y}[k] \right\rvert,
\label{eq:L1}
\end{equation}
for an output signal containing $K$ samples. 
If we assume that each processor conforms to the ideal model of a compressor, given here as the linear equivalent of Eq.~\eqref{eq:main_dB}:   
\begin{equation}
    y[k]=g[k]x[k],
\end{equation}
where $g[k] = 10^{\frac{g_{\mathrm{dB}}[k]}{20}}$, we can rewrite Eq.~\eqref{eq:L1} as
\begin{equation}\label{eq:L1_clean_comp}   
    L_1= \frac{1}{K}\sum_{k=1}^{K} \left\lvert x[k] \right\rvert \lvert g[k] - \hat{g}[k]\rvert.
\end{equation}
Under these conditions, $L_1$ represents the average sample-wise error between the gains, weighted by the magnitude of the input signal, making it a useful, if imperfect, proxy. However, its efficacy as a proxy for the error over $g[k]$ is diminished when the ideal compressor model is replaced by one that includes secondary effects:

\begin{equation}\label{eq:gain_hardware}   
    y[k] = g[k](h \circledast x)[k] + N[k],
\end{equation}
where $\circledast$ is the convolution operator, $N[k]$ is noise, and $h[k]$ is an impulse response that captures the linear effects of the signal chain outside of the broadband gain modulation imposed by the processor. These can include global delay and gain, as well as frequency-dependent group delay and amplitude changes imposed by anti-aliasing filters, buffering, and amplification stages. The $L_1$ error between an ideal digital processor and a hardware ground truth then becomes
\begin{equation}\label{eq:L1_hardware}   
    L_1 = \frac{1}{K}\sum_{k=1}^{K} \left\lvert g[k](h \circledast x)[k] + N[k] - \hat{g}[k]x[k] \right\rvert.
\end{equation}
Under these conditions, the per-sample differences between $(h \circledast x)[k] + N[k]$ and $x[k]$ can dominate the metric even when the effects of the filtering and noise are audibly imperceptible. 

Evaluating model performance in terms of $L_1$ over output waveforms can be particularly misleading when comparing models that synthesize the processed waveform directly to white- or gray-box signal-processing models, as the former may reproduce secondary effects introduced by the processor or capture chain—such as alterations to phase and/or magnitude response and the addition of noise—that the latter are rarely designed to simulate. This discrepancy is evident in \cite{gu2025solidstatebuscomplargescale}, where the smallest $L_1$ error among the black-box models is nearly an order of magnitude lower than that of the white-box models. When these same models are evaluated using multi-resolution STFT (M-STFT) error \cite{zhang2024amphion}, which compares short-time spectral magnitudes and is more robust to phase or temporal misalignment, the white-box model outperforms the black-box model. In this case, $L_1$ error over the output waveform primarily reflects the models’ ability to mimic secondary effects of the processor or artifacts of the capture chain, rather than their ability to replicate the gain behavior of the compressor.

The shortcomings of $L_1$ for quantifying the accuracy of dynamics processor models are further magnified when extracting $g[k]$ explicitly from the hardware input--output pairs. To eliminate division by zero and suppress numerical noise, we define a subset of sample indices that correspond to input signal values above a small threshold, $\epsilon$, given by
\begin{equation}\label{eq:extract_gain}
    \mathcal{K}_\epsilon = \{\, k : \lvert x[k] \rvert > \epsilon \,\}.
\end{equation}
For a digital compressor algorithm, the gain imposed by the model is given by
\begin{equation}
    \hat{g}[k] \approx \frac{\hat{y}[k]}{x[k]}, \quad k \in \mathcal{K}_\epsilon.
\end{equation}
However, applying the same operation to the hardware output in Eq.~\eqref{eq:gain_hardware} gives
\begin{equation}\label{eq:extract_gain_hardware}   
    \frac{y[k]}{x[k]}
    =
    \frac{g[k](h \circledast x)[k] + N[k]}{x[k]}
    \neq g[k].
\end{equation}
The gain-reduction signal is obscured by the noise created by sample-wise division, even under very modest perturbations of the signal outside the intended gain reduction. This limitation is not specific to $L_1$ over processed waveforms: when quantifying compressor model accuracy, any metric that computes loss per sample over $y[k]$ and $\hat{y}[k]$ is affected by the same sensitivity to secondary effects.

To sidestep the issues inherent with sample-wise waveform loss for this application, alternative metrics have been proposed. Wright and Välimäki \cite{wright2022grey} propose Multi-Resolution Short-Time Energy (MSTE) loss, in which the processed waveforms are pre-filtered to remove DC and then divided into \(M\) overlapping frames of length \(N_{f}\). The short-time energy (STE) loss is defined as the absolute difference in energy for each frame, averaged across all frames.
\begin{equation}
    \epsilon_{STE} = \frac{1}{M N_{f}}\sum_{m=0}^{M-1}\left|\sum_{n=0}^{N_{f}-1}\left( y_{p}[m,n]^{2} - \hat{y}_{p}[m,n]^{2}\right)\right|,
\end{equation}
where \(y_{p}\) and \(\hat{y}_{p}\) are the pre-filtered processed waveforms from the hardware and the model, respectively, $m$ indexes the frame, and $n$ indexes samples within each frame.

The MSTE loss is then obtained by evaluating \(\epsilon_{STE}\) at multiple values of \(N_{f}\), and taking the mean across these resolutions,
\begin{equation}
    \epsilon_{MSTE} = \frac{1}{I}\sum_{i=0}^{I-1} \epsilon_{STE,i}, 
\end{equation}
where \(I\) is the total number of frame lengths used.

While MSTE and related metrics help address the sensitivity of sample-wise metrics to secondary effects such as filtering and the addition of noise, they remain proxies for the underlying quantity of interest. In the context of model evaluation, metrics defined directly on the quantity of interest are preferable, as they represent an interpretable notion of error that is independent of the carrier waveform.

In the context of model training, a proxy metric is useful to the extent that minimizing the metric drives the model toward improved performance with respect to the true objective. A proxy can be functionally equivalent to the desired quantity, in the sense that it yields the same optimal model parameters, only insofar as it preserves the relative ordering of model error across all candidate solutions, for example when the proxy is a monotonic transformation of the true quantity.

Within this framework, MSTE constitutes a useful proxy, as it captures coarse agreement in signal energy across time. However, it is not guaranteed to be functionally equivalent to the underlying control signal for this application. MSTE measures agreement in frame-wise signal energy and may therefore be interpreted as operating on a short-time energy envelope estimate. As such, it is dependent on analysis parameters such as window length and overlap, and reflects a trade-off between temporal resolution and stability. Because envelope estimation from the waveform is not uniquely defined, different estimators or parameter choices may produce different results under identical conditions. Moreover, by aggregating energy within finite windows, MSTE is insensitive to the temporal distribution of energy within each frame, such that distinct gain trajectories can produce similar loss values.

The multi-resolution formulation mitigates these limitations by evaluating the metric at multiple time scales. However, the extent to which temporal discrepancies are resolved depends on the number and density of resolutions considered, and therefore on computational cost. Increasing the number of frame lengths increases the number of computations per training step, which can materially impact time to convergence in large-scale settings. Furthermore, no finite set of resolutions can guarantee equivalence with a model trained using loss computed directly on the quantity of interest.

In this work, we present an SSL bus compressor dataset consisting of input--output signals alongside the corresponding control voltage (CV) signal applied to the voltage-controlled amplifier (VCA). The CV signal included in the dataset is prescaled to correspond exactly with the gain reduction applied by the processor in decibels, $g_{\mathrm{dB}}[k]$. This enables users to process samples in this dataset using an existing model and, provided the model exposes its estimated gain trajectory, evaluate the performance of that model with respect to the principal quantity of the processor. The same measurements can be used to train new machine learning models using loss functions defined on the true objective, eliminating the need for proxy metrics altogether.

An additional advantage of capturing CV directly is that the control signal evolves on much slower time scales than the audio waveform. As a result, aggressive low-pass filtering can be applied without damaging the gain-reduction signal, yielding a measurement with substantially improved signal-to-noise ratio. By contrast, although the gain trajectory is low-frequency, it cannot be isolated from the audio by low-pass filtering without removing the carrier structure required to observe it.

\begin{figure}[ht]
\centerline{\includegraphics[width=\columnwidth]{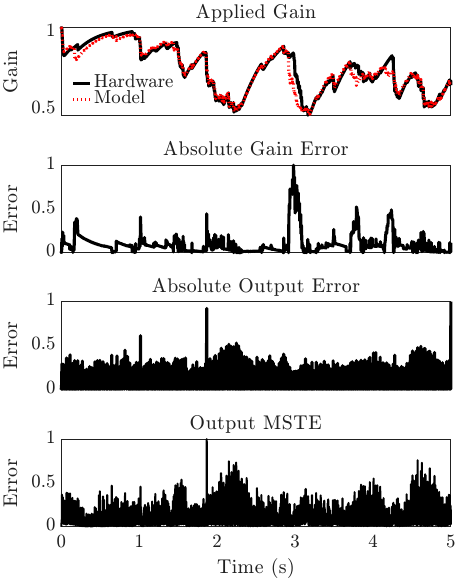}}
\caption{Dynamics processor error metrics. The top panel shows the measured linear gain signal, $g[k]$, from a hardware SSL compressor alongside the gain produced by a software model, $\hat{g}[k]$, for the same input. The second panel shows the absolute error between the linear gain signals, $|g[k] - \hat{g}[k]|$. The third panel shows the absolute error between the output waveforms, $|y[k] - \hat{y}[k]|$. The bottom panel shows the MSTE error between the output waveforms, where each frame’s contribution is distributed uniformly over the duration of that frame. The frame sizes used in the MSTE calculation are 8, 16, 32, and 64 samples, with 75\% overlap as suggested in \cite{wright2022grey}. All error signals are normalized by their respective maxima to facilitate comparison.}
\label{fig:error}
\end{figure}

The linear gain signals from a hardware SSL compressor and a digital model are shown in Fig.~\ref{fig:error}, along with sample-indexed representations of the three error metrics discussed in this paper. For the purposes of visualization, the error generated by each frame in the MSTE calculation was distributed uniformly over the samples in that frame, and the error shown at any sample is the summation of those contributions. The two proxy metrics, absolute error and MSTE over the output waveforms, are significantly noisier than the direct metric, absolute gain error. Additionally, the coarse structures of the proxy metrics track the trajectory of the direct metric only loosely and fail to capture salient structure in the gain-reduction error. This is particularly evident near the three-second mark, where there is a peak in gain-reduction error that is not captured by either of the proxy metrics.

\section{Experimental Evaluation}
\label{sec:experiment}
To compare each loss regime in context, an existing gray-box dynamics model, \texttt{torchcomp} \cite{yu2024differentiable}, was trained using each of the losses discussed above: $L_1$ over gain reduction, MSTE over processed waveforms, and $L_1$ over processed waveforms. These losses are denoted $L_1(g_{\mathrm{dB}},\hat{g}_{\mathrm{dB}})$, $\mathrm{MSTE}(y,\hat{y})$, and $L_1(y,\hat{y})$, respectively. 

To better align the model with the SSL topology, \texttt{torchcomp} was modified by replacing the RMS detector with a peak detector driven by the maximum instantaneous amplitude across channels. This produces a single shared gain control signal, which is applied identically to the left and right input channels. The optimized model parameters were threshold, ratio, attack time, release time, and makeup gain; all other aspects of the \texttt{torchcomp} compressor formulation were left unchanged.

It should be noted that there are aspects of the SSL behavior that this model cannot fully reproduce. \texttt{torchcomp} employs a static curve with an infinitely hard knee and a constant slope above the nominal threshold. While this is a reasonable approximation of the SSL behavior at the 10:1 ratio setting, its accuracy is diminished at lower ratios, where the transition into gain reduction becomes more gradual and the curvature of the static response above the threshold is more pronounced. Additionally, the ballistics filter in \texttt{torchcomp} is a fixed, single-pole design, while the \textsc{auto} release setting in the SSL compressor is implemented as a two-pole filter \cite{ssl_g384_schematics}, yielding program-dependent release trajectories. For this reason, the training data was restricted to examples measured with the ratio control set to 10:1 and single time constant release settings.

For each loss regime, ten independent models were trained, each using a single 30-second stereo excerpt drawn from the proposed dataset. The selected examples span five hardware control permutations, where each permutation denotes a distinct combination of compressor control settings. For each model, the checkpoint achieving the lowest value of its native training loss was retained. The training examples were selected pseudo-randomly from the eligible examples using a fixed seed, with at most one example drawn from any given song. The same set of ten examples was used across all loss regimes. For the two proxy metrics, prior to loss calculation at each step, the processed waveforms were pre-filtered to remove DC using the IIR filter proposed in \cite{wright2022grey}:
\begin{equation}
    H(z) = \frac{1-z^{-1}}{1-0.995z^{-1}}. 
\end{equation}

Each model was trained independently for 3000 steps using identical optimization settings. The number of steps was chosen empirically to ensure reliable convergence across all loss regimes. The Adam \cite{kingma2014adam} optimizer was employed with a learning rate of 0.01, and a ReduceLROnPlateau scheduler was applied with a factor of 0.5 and a patience of 25 epochs,  reducing the learning rate whenever the training loss failed to improve for 25 consecutive updates. All hyperparameters and model configurations were held constant across runs and loss regimes.

The purpose of this experiment is not to evaluate how this particular model generalizes, but rather to compare the extent to which different training objectives recover the underlying gain trajectory under otherwise identical optimization conditions. Accordingly, a traditional train-validation split is not used. Instead, each model is fit to a single measured example and evaluated on that same example in terms of each of the error metrics. This approach isolates the relationship between the optimized loss function and the control-signal objective: if a proxy loss is an effective surrogate for gain-reduction error, minimizing it should recover gain trajectories comparable to those obtained by minimizing gain-reduction error directly.

\begin{figure}[!t]
\centerline{\includegraphics[width=\columnwidth]{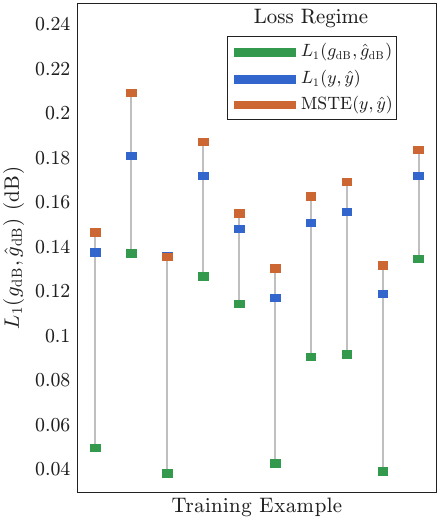}}
\caption{Gain-reduction signal error of trained models for each training example and loss regime. The gray lines connect the error values for the same training example.}
\label{fig:model_error}
\end{figure}

\definecolor{directmetric}{RGB}{255,244,232}

\begin{table}[!t]
\centering
\setlength{\tabcolsep}{3.5pt}
\renewcommand{\arraystretch}{1.05}
\caption{Training loss regime performance. For each training objective, the table reports mean error across ten models under the direct gain-reduction metric, $L_1(g_{\mathrm{dB}},\hat{g}_{\mathrm{dB}})$, and two waveform-domain proxy metrics. The waveform-domain metrics are computed after the DC-removal prefilter described above.}
\label{table:results}
{
\begin{tabular}{@{}l@{\hspace{4pt}}|@{\hspace{4pt}}ccc@{}}
\toprule 
\multicolumn{1}{@{}l}{} & \multicolumn{3}{c@{}}{Evaluation Metric} \\
\cmidrule(lr){2-4}
Training Loss
& $L_1(g_{\mathrm{dB}},\hat{g}_{\mathrm{dB}})$
& $L_1(y,\hat{y})$
& $\mathrm{MSTE}(y,\hat{y})$ \\[3pt]
$L_1(g_{\mathrm{dB}},\hat{g}_{\mathrm{dB}})$  
& \cellcolor{directmetric}\textbf{0.0869} \si{\decibel}
& $3.328{\times}10^{-3}$ 
& $2.746{\times}10^{-4}$ \\
$L_1(y,\hat{y})$            
& \cellcolor{directmetric}0.1493 \si{\decibel}
& $3.237{\times}10^{-3}$ 
& $2.444{\times}10^{-4}$ \\
$\mathrm{MSTE}(y,\hat{y})$  
& \cellcolor{directmetric}0.1615 \si{\decibel}
& $3.238{\times}10^{-3}$ 
& $2.423{\times}10^{-4}$ \\

\bottomrule
\end{tabular}
}
\end{table}

The $L_1$ gain-reduction error for models trained using each of the three loss regimes and each of the ten training examples is shown in Fig.~\ref{fig:model_error}. The mean error across the ten trained models, computed using $L_1(g_{\mathrm{dB}},\hat{g}_{\mathrm{dB}})$ and both proxy metrics, is presented in Table~\ref{table:results} for each training objective. As expected, each regime yields the lowest error under the metric for which it was optimized. However, the models are more clearly separated by gain-reduction error: the largest gain-reduction error is $85.9\%$ greater than the smallest gain-reduction error, whereas the corresponding error spreads are $2.8\%$ for $L_1(y,\hat{y})$ and $13.3\%$ for $\mathrm{MSTE}(y,\hat{y})$. This suggests that, under these conditions, direct gain-reduction supervision does not merely optimize an idiosyncratic metric at the expense of waveform-domain agreement; rather, evaluating the error through the processed waveform collapses some of the gain-trajectory structure made observable by the measured control signal, yielding similar waveform-domain errors for models that are clearly distinguishable under the direct metric. With respect to the central question of whether these metrics serve as effective proxies for gain-reduction error, the models trained using waveform-based losses do not achieve parity with the model trained directly on gain reduction when evaluated against the underlying control signal. This indicates that while the proxy losses are correlated with the underlying quantity of interest, minimizing them is not equivalent to minimizing gain-reduction error directly.

\section{Dataset}
\label{sec:dataset}
The dataset comprises 219 stereo input clips of 30 seconds each, sampled at \SI{44.1}{\kilo\hertz} with 24-bit resolution. For each clip, audio outputs were captured across 90 control permutations, yielding 19,710 stereo output examples. Each output is accompanied by a single-channel control signal sampled at \SI{44.1}{\kilo\hertz} with 32-bit resolution. Calibration signals and associated metadata are also included for each control permutation. The total dataset size is approximately 270~GB.

\subsection{Hardware}
\begin{figure}[ht]
\centerline{\includegraphics[width=\columnwidth]{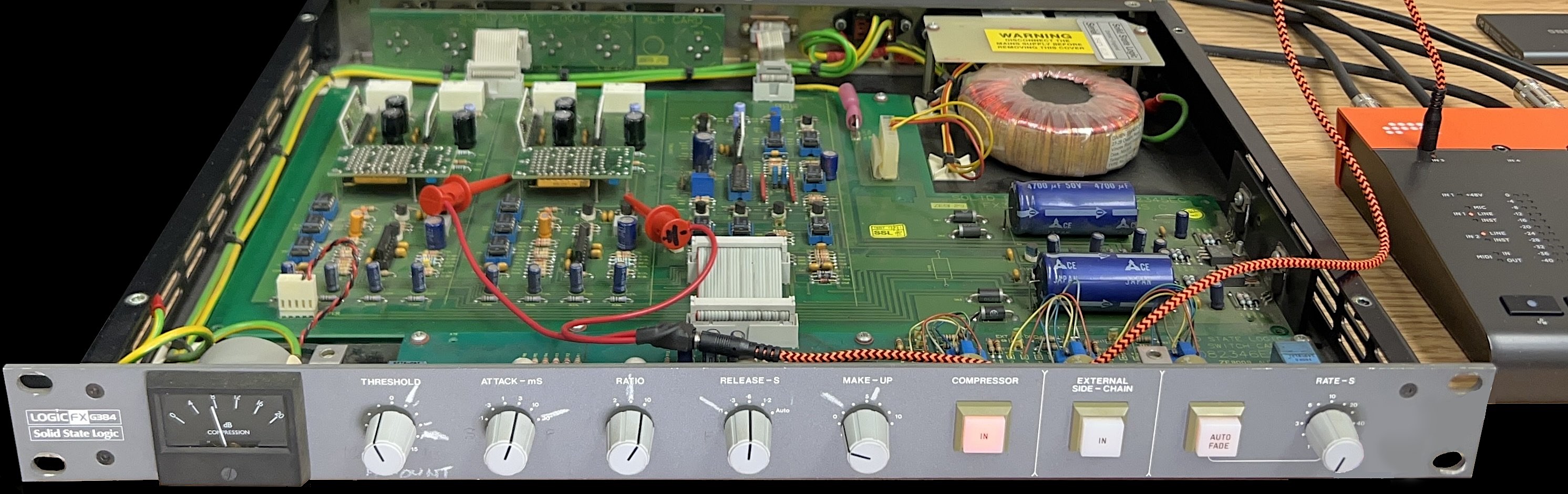}}
\caption{SSL Logic FX G384 bus compressor measured for dataset. The wire test leads are connected to the control input of the left audio VCA.}
\label{fig:hardware}
\end{figure}

First introduced as part of the center section of the Solid State Logic (SSL) 4000 B console in 1976, the SSL bus compressor has been produced in multiple revisions and form factors~\cite{ssl_bus_comp_guide}. The hardware unit measured for this dataset is a 1991 SSL Logic FX G384 bus compressor~\cite{ssl_g384_manual} shown in Fig.~\ref{fig:hardware}. The unit was fully serviced and calibrated prior to measurement. Additional trimmers were added at the control inputs of the audio VCAs to provide the same calibration flexibility available in the console version. The audio path VCAs in this unit are THAT Corporation 202-series devices~\cite{that202datasheet}, while the sidechain VCAs are DBX-branded 2151s~\cite{that2150datasheet}. As with other bus compressor derivatives, this design primarily differs from other versions of the SSL bus compressor in the input/output circuitry (balancing, debalancing, and line driver implementations) rather than the core dynamics control path~\cite{ssl_g384_schematics}.

\subsection{Input Signals}
\subsubsection{Music}
As the intended application of this class of compressor is the processing of a full stereo mix~\cite{ssl_gcomp500_userguide}, the most meaningful input signals are complete songs. Additionally, we require signals that have not already seen heavy dynamic range limiting. To meet these criteria, we use excerpts of the same input signals as in \cite{gu2025solidstatebuscomplargescale}, drawn from multitrack recordings accessed via the Cambridge MT search interface~\cite{cambridge_mt_search}. The shared input set provides an opportunity for direct comparison across datasets, isolating differences due to hardware implementation and control settings. 

For each source file containing \(N\) samples, one \SI{30}{\second} excerpt was selected by drawing the starting sample index \(k_0\) pseudorandomly from the viable range
\begin{equation}
0 \le k_0 \le N - 30f_s,
\end{equation}
where \(f_s\) is the sample rate. Files with other sample rates were resampled to \SI{44.1}{\kilo\hertz}, and each excerpt was given \SI{10}{\milli\second} fades at its boundaries before peak normalization to a maximum absolute sample value of 0.9. The music inputs are comprised of 219, \SI{30}{\second} stereo clips sampled at \SI{44.1}{\kilo\hertz} and stored with 24-bit resolution.

\subsubsection{Calibration Signals}\label{sect:cal}
In addition to the musical input signals, we include several calibration signals processed at each control permutation. These include a slow ramp with an increase in level that is linear in decibels. The ramp consists of a \SI{1}{\kilo\hertz} sine-wave carrier, the amplitude of which rises from \SI{-48}{\decibel} to \SI{0}{\decibel} at a rate of \SI{1.6}{\decibel\per\second}. From the compressor’s CV response to this input, it is possible to approximate the processor’s static curve, provided that the selected attack time is fast relative to the rate of increase of the ramp. Under this condition, the compressor operates in a quasi-static regime, allowing the static input--output relationship to be observed directly. A functional threshold can likewise be identified from the point at which the CV departs from its baseline.

Three burst signals of different durations are also included. As with the ramp, these are constructed from \SI{1}{\kilo\hertz} sine-wave carriers and are designed to cross the threshold and drive the compressor into gain reduction. The compressor’s CV response to a burst captures its attack and release behavior. In the case of the \textsc{auto} release setting, the dependence of the response on burst duration illustrates the program-dependent release mechanism.

The final calibration signal included for each control permutation is a maximum-length sequence with amplitude chosen to remain below the threshold. Because this signal does not engage the gain reduction mechanism, it can be used in conjunction with the corresponding audio output to determine the relative delay between input and output signals, enabling precise time alignment where required.

We also include a global set of calibration signals for use in determining the scaling factor that translates the measured CV signal to gain reduction in decibels. This set is comprised of eleven discrete bursts with amplitudes increasing in one-decibel steps from \SI{4}{\decibel} to \SI{14}{\decibel} above the threshold, with each burst long enough to allow the ballistics filter to reach steady state so that the static response to that input can be recorded.

To estimate the scaling factor, the responses are trimmed to the static region, gain reduction in decibels is computed from the audio input--output relationship, and a linear model of the form $g_{\mathrm{dB}} = a \cdot \mathrm{CV}$ is fit to the measured data. The slope, $a$, of this fit defines the conversion factor from CV to gain reduction in decibels.

Note that this factor has already been applied to the measured CV signals in the dataset, so the provided CV signals are expressed directly as gain reduction in decibels. The calibration signals are included for completeness and to allow end users to reproduce the scaling factor calculation if desired.

\subsection{Control Permutations}

The relevant user-facing controls on the Logic FX G384 compressor, along with the range of settings for each, are shown in Table~\ref{table:controls}. To balance coverage with dataset size, all permutations of the switched controls were included for a single threshold and makeup gain setting. Multiple settings for these parameters were not included, as neither represents an independent degree of freedom. 

In the hardware unit, makeup gain is implemented by applying a constant DC offset to the control signal that determines the gain of the audio VCA, while leaving the gain of the sidechain VCA unchanged. This results in a simple scaling of the output waveform. As the control signal is supplied in \si{\decibel}, additional permutations can be synthesized by applying the desired change in makeup gain as an offset in \si{\decibel} to both the control signal and the output audio.

Similarly, the threshold parameter enters the gain law only as an additive offset in the dB-domain sidechain VCA control signal. Applying a constant offset $\Delta$ to the input signal in decibels,
\begin{equation}
x_{\mathrm{dB}}[k] \mapsto x_{\mathrm{dB}}[k] + \Delta,
\end{equation}
is equivalent to replacing the threshold parameter with an effective threshold,
\begin{equation}
T_{\mathrm{eff}} = T - \Delta,
\end{equation}
leaving the control trajectory $g_{\mathrm{dB}}[k]$ unchanged. To preserve consistency of the input--output pair, the same offset must also be applied to the audio output signal,
\begin{equation}
y_{\mathrm{dB}}[k] \mapsto y_{\mathrm{dB}}[k] + \Delta.
\end{equation}
Accordingly, additional threshold conditions can be synthesized by applying a common gain offset to both the input and output audio signals while leaving the control signal unchanged. 

\begin{table}[t]
\tabcolsep6pt
\caption{Control parameters and available settings.}
\label{table:controls}
{%
\begin{tabular}{@{}lllc@{}}
\toprule
Control & Resolution & Unit & Settings\\
\midrule
Threshold   & Continuous & \si{\decibel}       & $[-15,\,15]$ \\
Attack      & Switch     & \si{\milli\second}  & 0.1, 0.3, 1, 3, 10, 30 \\
Ratio       & Switch     & dB/dB               & 2, 4, 10 \\
Release     & Switch     & \si{\second}        & 0.1, 0.3, 0.6, 1.2, Auto \\
Makeup Gain & Continuous & \si{\decibel}       & $[-5,\,15]$ \\
\bottomrule
\end{tabular}}
\end{table}
\subsection{Captured Data}
The responses to each input were captured using a Bitwig Connect 4/12 audio interface \cite{bitwig_connect412_userguide} at a sample rate of \SI{44.1}{\kilo\hertz}. This interface includes the DC-coupled inputs necessary to capture CV signals. 

Stereo audio outputs were stored at a resolution of 24 bits per sample. The raw captured audio output reflects any gain applied by the gain cell in the processor along with any additional gain in the signal chain. To simplify interpretation, any static input--output gain that is not explained by the CV signal was removed prior to storage. The factor by which it was adjusted is stored in the metadata, allowing recovery of the original signal if required.

The stored CV is scaled so that it directly represents gain reduction in \si{\decibel}, as outlined in subsection \ref{sect:cal}. Since the signal controlling both audio VCAs is identical up to a small calibration scalar, the CV from a single VCA was captured. The CV signals were stored at a resolution of 32 bits per sample to accommodate the large amplitudes expected. Note that while the CV signals were saved in the WAV format, they are not audio signals; they can contain large DC components and are not restricted to the range $[-1,\,1]$. Appropriate caution should be exercised such that these signals are not routed to amplifiers or loudspeakers. 

\subsection{Temporal Alignment of Measured Signals}

At capture, care was taken to measure and compensate for system latency. As a result, the input and output audio signals are aligned in time to within one sample.

All signals within a given control permutation were captured using a single instance of a MATLAB \texttt{audiostreamer} object, so any residual capture offset is consistent across the recorded channels for that permutation~\cite{mathworks_audiostreamer}. Sub-threshold MLS calibration signals are provided for each permutation to enable users to verify or refine audio signal alignment if desired.

A procedure is provided here for confirming or refining the alignment of the control signal to the audio signals, should that be necessary for a user's application. The measured CV is converted to a linear gain, and the integer lag that minimizes the discrepancy between the measured output and the gain-scaled input is selected. Specifically, the estimated lag $\ell^\ast$ is given by
\begin{equation}
\ell^\ast = \argmin_{-L \le \ell \le L}
\frac{1}{|\mathcal{K}_\ell|}
\sum_{k \in \mathcal{K}_\ell}
\left| y(k) - g(k+\ell)x(k) \right|,
\end{equation}
where $L$ defines the maximum lag considered in samples, and $\mathcal{K}_\ell$ denotes the set of indices for which the shifted signals overlap. After estimating $\ell^\ast$, the control signal can be shifted accordingly. This approach can be extended to estimate fractional-sample offsets by allowing non-integer lags and evaluating the objective under interpolation.

\subsection{Metadata}

In addition to the control settings described in Table~\ref{table:controls}, several derived quantities are included in the dataset metadata to facilitate interpretation and reuse.

In the SSL topology, the effective threshold of the processor is determined by a combination of the offset voltage applied to the sidechain VCA through the threshold potentiometer and an offset current applied at the slope amplification stage. Since the value of the current offset is tied to the position of the ratio switch, the input level at which compression begins is not determined solely by the threshold control.

A derived effective threshold is provided in units of \si{\decibel}, defined operationally from the compressor response to the ramp calibration signal described in Section~\ref{sect:cal}. The control signal is analyzed in a region below the threshold to estimate its baseline and variance, and the threshold is taken as the input level at which the control signal first exceeds this baseline by five standard deviations. This response-defined threshold corresponds to the onset of observable gain reduction in the measured unit. Because the onset of compression is gradual, particularly at lower ratios, this quantity should not be interpreted as equivalent to the threshold parameter of an idealized static-curve model. Users interested in parameter recovery may therefore wish to transform or redefine the threshold according to the parameterization of the model under study.

For reproducibility and completeness, the nominal threshold setting is also included in the metadata. Because the front-panel markings associated with the threshold control are sparse, this quantity is reported as the measured voltage at the wiper of the potentiometer.

For the same reason, the makeup gain setting is likewise reported as the measured voltage at the wiper of its corresponding potentiometer. A second makeup gain quantity is also included in the metadata that is derived from the CV signal when the processor is not actively in gain reduction. It is this second value that corresponds to the actual static gain applied by the processor in \si{\decibel}.

In addition to the nominal attack time and release time settings, the corresponding component values in the ballistics filter are included along with the associated time constants. These quantities may be used in the context of a circuit-informed model to predict the expected ballistics behavior and to verify model outputs in a parameter recovery experiment. It should be noted, however, that the effective attack times are greatly accelerated by the loop gain of the control circuit. Additionally, because loop gain is not constant, the attack trajectory deviates from an ideal exponential response.

\section{Conclusions}
\label{sec:conclusions}
In this work, we identified the gain control signal, $g_{\mathrm{dB}}[k]$, as the primary quantity of interest when evaluating the accuracy of a DRC model's gain-reduction behavior. In existing DRC datasets, this quantity is not directly observable from the included input--output signals, and models are trained using proxy metrics computed over the processed waveforms. We characterized the limitations of common proxy loss regimes, including $L_1$ and MSTE computed over the processed audio, which can be influenced by secondary effects such as phase shifts and noise, even when these effects are perceptually insignificant. To evaluate these metrics in context, we trained a differentiable compressor model under each loss regime. When evaluated with respect to the ground-truth gain trajectory, models trained using the proxy losses did not achieve parity with the model trained by minimizing error in $g_{\mathrm{dB}}[k]$ directly, indicating that they are not equivalent surrogates for gain-reduction error in this setting. Additionally, the waveform-domain metrics assigned comparatively similar errors to models that were clearly separated by the direct gain-reduction metric, suggesting that these proxy objectives can obscure some of the salient processor behavior.

We also introduced a dataset of measured music and calibration signals processed by a hardware SSL bus compressor that includes the corresponding gain-reduction signal, $g_{\mathrm{dB}}[k]$, in the form of CV measured alongside the audio signal, enabling direct supervision of the gain reduction and allowing proxy metrics to be evaluated against a ground-truth measure of error. Future work will extend this approach to additional hardware processors and explore models that capture behavior specific to the SSL compressor that generic models cannot replicate.

\section{Acknowledgments}
\redact{The authors thank Ric Ocasek for bequeathing the hardware used in this work to the University of Rochester Audio and Music Engineering program. This work was supported by a Research Initiation Award from the SoundSpace Institute.} 

\bibliographystyle{IEEEtranDAFx}
\bibliography{DAFx26_tmpl} 

\end{document}